\documentclass[twocolumn,preprint,amsmath,amssymb,amsfonts,3p]{elsarticle}

\usepackage{float} 

\usepackage{comment}
\usepackage[normalem]{ulem}
\usepackage[english]{babel}
\usepackage{graphicx}
\usepackage{dcolumn}
\usepackage{bm}
\usepackage{blindtext}
\usepackage{slashed}
\usepackage{verbatim}
\usepackage{mathrsfs}
\usepackage{amsmath}
\usepackage{cancel}
\usepackage{physics}
\usepackage{epstopdf}
\usepackage{mathtools}
\usepackage{color}
\usepackage{physics}
\usepackage{tensor}

\newcommand{\be}{\begin{equation}}
\newcommand{\ee}{\end{equation}}
\newcommand{\ba}{\begin{eqnarray}}
\newcommand{\ea}{\end{eqnarray}}

\usepackage[usenames,dvipsnames]{pstricks}
\usepackage{epsfig}
\usepackage{pst-grad} 
\usepackage{pst-plot} 

\usepackage{hyperref}

\usepackage{verbatim}

\begin{document}

\title{Taub-NUT solutions in conformal electrodynamics}

\author[1,2]{Alvaro Ballon Bordo}
\ead{aballonbordo@perimeterinstitute.ca}

\author[1,2]{David Kubiz{\v n}\'ak}
\ead{dkubiznak@perimeterinstitute.ca}

\author[1,2]{Tales Rick Perche}
\ead{trickperche@perimeterinstitute.ca}

\address[1]{Perimeter Institute for Theoretical Physics, Waterloo, Ontario, N2L 2Y5, Canada}
\address[2]{Department of Physics and Astronomy, University of Waterloo, Waterloo, Ontario, Canada, N2L 3G1}

\begin{abstract}
We construct a novel charged Taub-NUT spacetime, 
providing a first non-trivial example of a self-gravitating solution to the recently proposed ModMax theory \cite{Bandos:2020jsw, Kosyakov:2020wxv}, the most general (1-parametric) theory of non-linear electrodynamics that is characterized by both the conformal symmetry and the $SO(2)$ duality-rotation invariance. 
The spacetime features non-trivial magnetic fields, in their presence the non-linearity of the field becomes apparent and the solution is distinguished from that of the Maxwell theory. Thermodynamics of the new solution, the charged black hole limit of vanishing NUT parameter, and the self-dual solutions are also briefly discussed.

\end{abstract}


\date{23 March, 2021}

\maketitle

\section{Introduction}

Maxwell's electromagnetism is a relativistic theory of linear electrodynamics which enjoys many fundamental  symmetries, the two distinguished being the conformal invariance and the invariance under the $SO(2)$ duality-rotation transformations. Over the years,
various non-linear extensions of this theory that modify the behavior of the field in a strong field regime have been considered. Perhaps the best known example of a theory of {\em non-linear electrodynamics} is the 
Born--Infeld theory \cite{born1934foundations}. Although invariant under duality transformations \cite{gibbons1995electric}, the Born--Infeld theory contains a dimensionful parameter, and is not conformally invariant.

Remarkably, Maxwell's theory is not the only (four-dimensional) theory that enjoys conformal and duality-rotation symmetries. 
As shown recently \cite{Bandos:2020jsw, Kosyakov:2020wxv}, there is a 1-parametric generalization of the Maxwell theory with these properties. 
To stress its conformal invariance, in this paper we call this theory {\em conformal} (non-linear) {\em electrodynamics}. The corresponding Lagrangian density reads \cite{Bandos:2020jsw, Kosyakov:2020wxv}
\be\label{NLELag}
    \mathcal{L} = -\frac{1}{2} \left({\cal S}\cosh\gamma - \sqrt{{\cal S}^2+{\cal P}^2}\sinh\gamma\right)\,,
\end{equation}
where $\gamma$ is a free dimensionless parameter, and ${\cal S}$ and ${\cal P}$ are the two invariants of the electromagnetic field,
\be
{\cal S}=\frac{1}{2}F_{\mu\nu} F^{\mu\nu}\,,\quad {\cal P}=\frac{1}{2}F_{\mu\nu} (*F)^{\mu\nu}\,, 
\ee
with the field strength $F_{\mu\nu}$ given by 
$F_{\mu\nu}=\partial_\mu A_\nu-\partial_\nu A_\mu$, in terms of the vector potential 
$A_\mu$.  For $\gamma=0$, the theory reduces to the linear Maxwell theory. As discussed in \cite{Bandos:2020jsw} when $\gamma\neq 0$ a birefringe phenomenon occurs; apart from the light-like polarization mode there is another mode which is subluminal for $\gamma>0$ and superluminal for $\gamma<0$, hinting on a physical restriction $\gamma\geq 0$.  

It is the aim of this paper to construct a first non-trivial self-gravitating solution of the conformal electrodynamics. As can be seen from the Lagrangian \eqref{NLELag}, the non-linear character of the theory only appears for solutions with magnetic fields for which the invariant ${\cal P}$ is non-trivial. In what follows, we construct `the simplest possible' such solution --- the charged Taub-NUT(-AdS) spacetime. Even in the absence of magnetic charges, this spacetime features charged Misner strings and associated with them magnetic fields. In their presence, the invariant ${\cal P}$ becomes non-trivial, the non-linearity of the field becomes apparent, and the solution is distinguished from that of the Maxwell theory.

The Taub--NUT spacetime \cite{taub1951empty, newman1963empty} is a remarkable solution of the Einstein equations that has been a subject of controversy for many years, e.g. \cite{misner1963flatter, misner1967taub}.    
Although predominantly studied in the Euclidean setting (see e.g. recent \cite{Ciambelli:2020qny} and references therein),
it also provides an example of a cosmological model and a black hole spacetime. In what follows we concentrate on the (Lorentzian) black hole case, maintaining the
Misner string singularities on the axes \cite{bonnor1969new, miller1971taub, Manko:2005nm, Clement:2015cxa, Clement:2015aka,Durka:2019ajz, 
Kubiznak:2019yiu,  Ballon:2019uha, Bordo:2019tyh, Bordo:2019rhu, Clement:2019ghi, Bordo:2020kxm,
Kalamakis:2020aaj}. Such a spacetime is geodesically complete and (despite the presence of regions with closed timelike curves) it is free of causal pathologies for geodesic observers \cite{miller1971taub, Clement:2015cxa, Clement:2015aka}. Moreover, as shown recently, the corresponding black hole thermodynamics can be understood by standard 
thermodynamic techniques \cite{Durka:2019ajz, Bordo:2019tyh,Kubiznak:2019yiu, Clement:2019ghi} and yields a consistent first law,  even in the presence of rotation and charges \cite{Ballon:2019uha, Bordo:2019rhu, Bordo:2020kxm}.

Our paper is organized as follows. In the next section we review the basic properties of the conformal electrodynamics. In Sec.~\ref{secNUT} we construct the charged Taub-NUT(-AdS) spacetime in this theory and compare it briefly with its Maxwell counterpart. 
Sec.~\ref{secThermo} is devoted to the study of the thermodynamics of the obtained solution, while Sec.~\ref{n0} discusses the charged black hole limit of vanishing NUT charge. We conclude in Sec.~\ref{secConclusion}.

\section{Conformal electrodynamics}\label{secNLE}

As shown in \cite{Bandos:2020jsw, Kosyakov:2020wxv}
the theory \eqref{NLELag} is the most 
general four-dimensional electrodynamics that possesses both the conformal symmetry and  the $SO(2)$ duality-rotation invariance. 
Introducing the following  `material' field strength tensor: 
\be
E_{\mu\nu} = \pdv{\mathcal{L}}{F^{\mu\nu}}
=2\Bigl({\cal L_S}F_{\mu\nu}+{\cal L_P}*\!F_{\mu\nu}\Bigr)\,, 
\ee
where 
\ba
{\cal L_S}&=& \pdv{\mathcal{L}}{\mathcal{S}}
=\frac{1}{2}\Bigl(\frac{\cal S}{\sqrt{\mathcal{S}^2 + \mathcal{P}^2}}\sinh \gamma\nonumber\\
&&-\cosh\gamma\Bigr)\,,\\
{\cal L_P}&=&\pdv{\mathcal{L}}{\mathcal{P}}
=\frac{1}{2}\frac{\cal P}{\sqrt{\mathcal{S}^2 + \mathcal{P}^2}}\sinh \gamma\,,
\ea
the field equations are simply written as 
\be\label{FE}
d*E=0\,,\quad
dF=0\,, 
\ee
and are invariant under the $SO(2)$ duality rotations 
\begin{equation}\label{dualityRot}
    \begin{pmatrix}
    E'_{\mu\nu}\\
     *F'_{\mu\nu}
    \end{pmatrix}
    =
    \begin{pmatrix}
        \cos\theta & \sin \theta\\
        -\sin\theta & \cos\theta 
    \end{pmatrix}
    \begin{pmatrix}
        E_{\mu\nu}\\
        *F_{\mu\nu}
    \end{pmatrix}.
\end{equation}

Similarly, under a conformal transformation of the metric, $g\to \Omega^2 g$, we find 
\be 
F\to F\,,\quad *F\to *F\,,
\ee
and since  ${\cal S}\to \Omega^{-4} {\cal S}, 
{\cal P}\to \Omega^{-4}{\cal P}$, and 
${\cal L_S}\to {\cal L_S}, {\cal L_P}\to {\cal L_P}$, we also have 
\be 
E\to E\,,\quad *E\to *E\,,
\ee 
and the field equations \eqref{FE} are invariant 
under conformal transformations.

The electromagnetic stress-energy tensor reads
\be
T^{\mu\nu}=-\frac{1}{4\pi}\Bigl(2F^{\mu\sigma}F^{\nu}{}_\sigma {\cal L_S}+{\cal P}{\cal L_P} g^{\mu\nu}-{\cal L}g^{\mu\nu}\Bigr)\,,
\ee
and, using the fact that 
\be\label{trace} 
{\cal L_S S}+{\cal L_P P}={\cal L}\,,
\ee
it is obviously traceless, 
$
T^\mu{}_\mu=0\,.
$
Employing \eqref{trace}, the energy momentum tensor can be further recast as
\be\label{Tmunu}
T^{\mu\nu}=\frac{1}{4\pi}\Bigl({\cal S} g^{\mu\nu}-2F^{\mu\sigma}F^{\nu}{}_\sigma \Bigr){\cal L_S}\,.
\ee

In what follows we shall calculate the 
electric and magnetic charges inside a closed spacelike two-surface $\mathcal{S}$. These are simply given by 
\be\label{charges}
q_e = \frac{1}{4\pi} \int_{\mathcal{S}}* E\,, \quad q_m = \frac{1}{4\pi}\int_{\mathcal{S}} F\,.
\ee
These definitions respect the duality rotations of Eq. \eqref{dualityRot}, in the sense that magnetic charges rotate to electric charges and vice-versa.

\section{Charged Taub-NUT solution}
\label{secNUT}

Let us now couple the conformal electrodynamics 
to gravity and construct the corresponding (spherical topology) charged Taub-NUT(-AdS) solution. We thus consider the following bulk action for the coupled theory with gravity: 
\begin{equation}\label{bulkAct}
    I= \frac{1}{16\pi} \int_{M}\dd^4 x \sqrt{-g}\left(R -2\Lambda\right)-\frac{1}{4\pi}\int \dd^4 x\sqrt{-g}\mathcal{L}\,.
\end{equation}
Here, $\Lambda$ stands for the negative cosmological constant $\Lambda=-3/l^2$.

We seek the Taub-NUT solution characterized by a single metric function $f=f(r)$,  
\ba\label{ansatz}
    ds^2 &=&{ -f\big(dt + 2n \cos\theta d \phi\big)^2}+\frac{dr^2}{f} \nonumber\\
 &&\quad + (r^2+n^2)\left(d\theta^2 +\sin^2\!\theta d\phi^2\right)\,,\ 
\ea
and the following gauge potential:
\be\label{A}
A = a(dt + 2n\cos\theta d\phi)\,, 
\ee
where the gauge function $a=a(r)$. In the above, we  have denoted the NUT parameter by $n$, and chosen a symmetric distribution for Misner strings, which are now located on both the north-pole and south-pole axes.

With this ansatz we find the following expressions for 
$F$ and $*F$:
\ba 
    F &=& -a' dt\wedge dr + 2na'\cos\theta dr\wedge d\phi \nonumber\\
    &&- 2n a\sin\theta d\theta\wedge d\phi\,,\\
*F &=& {-}\frac{2n a}{n^2+r^2}dt\wedge dr {+}\frac{4n^2a}{n^2 + r^2} \cos\theta dr\wedge d\phi\nonumber\\
&& {+}(n^2 + r^2)a' \sin\theta d\theta \wedge d\phi\,.
\ea
These are independent of the metric function $f$ on behalf of the fact that the determinant of the metric \eqref{ansatz} simply reads $\sqrt{-g}=r^2+n^2$. 
The two electromagnetic invariants take a rather elegant shape
\be
    {\cal S}= -{  a'^2}+\frac{4 n^2 a^2}{(n^2+r^2)^2}\,,\quad {\cal P} = {-}\frac{4 n a a'}{n^2+r^2}\,,
\ee
and so {do the material tensors}
\ba
 E \!\!&=&\!\! {a' e^{\gamma}}dt\wedge dr {-} 2n a'e^\gamma \cos\theta dr\wedge d\phi\nonumber\\
    \!\!&&\!\! \quad {+}2n a e^{-\gamma} \sin\theta d\theta \wedge d\phi\,,\\
    *E \!\!&=&\!\! {\frac{2an e^{-\gamma}}{n^2+r^2} dt\wedge dr -\frac{4n^2a e^{-\gamma}}{n^2+r^2}\cos\theta dr\wedge d\phi}\nonumber\\
    \!\!&&\!\! \quad {-(n^2+r^2)a'e^\gamma \sin\theta} d\theta \wedge d\phi\,.
\ea
The only non-zero component of the field equation \eqref{FE} then yields 
\begin{align}
-e^{\gamma } \Bigl[\left(n^2+r^2\right) a''+2 r a'\Bigr]-\frac{4 e^{-\gamma } n^2 a}{n^2+r^2}=0\,,
\end{align}
which when integrated gives 
\begin{equation}
    a = c_1 \sin\left(2e^{-\gamma} \arctan\frac{r}{n}\right)+c_2 \cos\left(2e^{-\gamma} \arctan\frac{r}{n}\right),
\end{equation}
where the integration constants $c_1$ and $c_2$ are  related to the electric and magnetic charges of the black hole, as described below.

Having obtained the gauge potential, let us now turn towards the Einstein equations with the electromagnetic tensor given by \eqref{Tmunu}. They yield the following equation for the metric function $f$: 
\begin{align}
    \frac{3}{l^2} + \frac{f(n^2-r^2)+(n^2+r^2)(r f'-1)}{(n^2+r^2)^2}-e^\gamma a'^2\nonumber\\
    -\frac{4 e^{-\gamma}n^2 a^2}{(n^2+r^2)^2}=0\,,
\end{align}
giving the following solution:
\be 
    f = \frac{ 4e^{-\gamma}n^2(c_1^2\!+\!c_2^2)\!-\! 2mr+r^2-n^2}{r^2+n^2}-\frac{ 3 n^4-6n^2 r^2-r^4}{l^2(r^2+n^2)}\,,
\ee
where the integration constant $m$ represents the asymptotic mass of the solution (as can be seen by a large $r$ expansion). 

The final step is to determine the constants $c_1$ and $c_2$ in terms of the asymptotic electric and magnetic charges. To this purpose we employ formulae \eqref{charges}, choosing $S$ to be a sphere of radius $r$. Similar to the Maxwell case, the enclosed charges are radius dependent: 
\be\label{QeQm}
    q_e = e^\gamma (n^2+r^2)a'\,,\quad q_m = 2n a\,,
\ee
reflecting the fact that the Misner strings carry electromagnetic charges. Evaluating these asymptotically, it is usual to define the asymptotic value of the electric and magnetic charges as the electric and magnetic parameters $e$ and $g$,
\be
\lim_{r\to\infty} q_e = e\,,\quad \lim_{r\to\infty} q_m = -2n g\,.
\ee
In terms of these, the constants $c_1$ and $c_2$ above are determined as follows: 
\begin{align}
    c_1 &= -g \cos\left(2 e^{-\gamma}\pi\right)- \frac{ e \cos(2 e^{-\gamma}\pi)}{2n}\,,\\
    c_2 &= -g \sin\left(2 e^{-\gamma}\pi\right)+ \frac{ e \sin(2 e^{-\gamma}\pi)}{2n}\,.
\end{align}

To summarize, the charged Taub-NUT(-AdS) solution constructed in this section takes the form \eqref{ansatz}, \eqref{A}, where the metric and gauge functions are given by 
\begin{equation}
    f = \frac{ e^{-\gamma}(e^2\!+\!4n^2g^2)\!-\! 2mr\!+\!r^2\!-\!n^2}{r^2+n^2}-\frac{ 3 n^4\!-\!6n^2 r^2\!-\!r^4}{l^2(r^2+n^2)}\,,
\end{equation}
and 
\begin{align}
    a =& -g\cos\Bigl[ e^{-\gamma}\left(\pi-2\arctan\frac{r}{n}\right)\Bigr]\nonumber\\&- \frac{e}{2n} \sin\Bigl[ e^{-\gamma} \left(\pi - 2\arctan\frac{r}{n}\right)\Bigr]\,.
\end{align}

To compare with the charged Taub-NUT solution in the Maxwell theory, we see that at the level of the metric function $f$, the effect of considering the conformal electrodynamics rather than the Maxwell theory corresponds to rescaling the electric and magnetic parameters by $e^{-\gamma/2}$.
To compare the gauge functions, we expand $a$ as a power series in $\gamma$: 
\ba 
a &=& g \frac{r^2-n^2}{r^2 + n^2} + \frac{er}{r^2+n^2}\nonumber\\
&&+ \gamma\left(\frac{e(r^2-n^2)}{2n(n^2 + r^2)} - \frac{2gnr^2}{(n^2 + r^2)}\right)\nonumber\\
&& \times \Bigl(\pi - 2\arctan\frac{r}{n}\Bigr)
    +\mathcal{O}(\gamma^2)\,.
\ea
The leading term corresponds to the solution in the Maxwell theory, while the next term represents a  non-linear effect that arises from the theory \eqref{NLELag} to first-order in $\gamma$.

\section{Thermodynamics}\label{secThermo}
In this section we discuss some basic properties of the obtained solution and its thermodynamics {(see \cite{Bokulic:2021dtz} for a general discussion of black hole thermodynamics in the presence of non-linear electromagnetic fields).} We follow the Lorentzian approach developed in \cite{Durka:2019ajz, 
Kubiznak:2019yiu,  Ballon:2019uha, Bordo:2019tyh, Bordo:2019rhu, Clement:2019ghi, Bordo:2020kxm}. Namely, we maintain the Misner strings and assign to them physical properties such as temperature/angular velocity. In this way we are able to formulate a full cohomogeneity first law where all the physical parameters of the solution are varied 
independently.\footnote{In this paper we focus on a configuration with symmetric distribution of Misner strings and do not consider the `string strength' as a  thermodynamic quantity. Our analysis can be straightforwardly extended to this more general case by following 
\cite{Bordo:2019tyh}.} The results of this section can be straightforwardly compared to those for charged Taub-NUT(-AdS) spacetimes in Maxwell's theory  \cite{Ballon:2019uha}. 

\subsection{Basic thermodynamic quantities}
To start our discussion we first turn to the black hole horizon. This is located at the largest root $r_+$ of $f(r_+)=0$. It is a Killing horizon generated by the Killing vector 
\be
k=\partial_t\,. 
\ee
In what follows we identify the temperature of the spacetime with the black hole horizon temperature: 
\begin{equation}
    T = \frac{f'(r_+)}{4\pi} = \frac{1}{4\pi r_+}\left(1+\frac{3(r^2_+ + n^2)}{l^2} - e^{-\gamma} \frac{e^2 + 4n g^2}{r_+^2+n^2}\right)\,,
\end{equation}
and the entropy of the spacetime with the area of the black hole horizon: 
\begin{equation}
    S = \frac{\mbox{Area}}{4} = \pi(r_+^2+n^2)\,.
\end{equation}

In addition to the black hole horizon, there are two additional Killing horizons in the spacetime that are associated with the Misner strings. They are generated by the following Killing vectors:
\be\label{kpm}
k_\pm=\partial_t \mp \frac{1}{2n} \partial_\phi\,. 
\ee
Contrary to the vector $k$, these Killing vectors are not properly normalized at infinity. In what follows we identify the associated surface gravity $\kappa_\pm$ with the `Misner potential' $\psi$:
\be
\psi=\frac{\kappa_\pm}{4\pi}=\frac{1}{8\pi n}\,. 
\ee 
Alternatively \cite{Durka:2019ajz,Clement:2019ghi}, $\psi$ can be attributed a meaning of the angular velocity of the string, c.f. \eqref{kpm}. We call the conjugate quantity to $\psi$ the Misner charge and denote it by $N$.

Similar to the case of Maxwell electrodynamics, the  
thermodynamic mass $M$ is simply given by the  parameter $m$,
\be
M=m\,, 
\ee
and the asymptotic angular momentum of the spacetime vanishes. 
 
Since the Misner strings are charged, the electric and magnetic charges depend on the radius of the sphere and are given by \eqref{QeQm}. 
One can easily check that they are related, $q_e\leftrightarrow q_m$, by the electromagnetic duality:
\be\label{duality}
e \leftrightarrow -2ng\,,\quad 2ng  \leftrightarrow e\,.
\ee
In particular, we have 
\be
Q=e\,,\quad Q_m=-2ng\, 
\ee 
for the asymptotic charges, and 
\ba\label{horizoncharges}
Q^+&=&q_e(r_+)= e^\gamma (n^2+r_+^2)a'(r_+)\,,\nonumber\\
Q_m^+&=&q_m(r_+)=2na(r_+)\,
\ea
for the horizon charges. 
The electrostatic potential $\phi$ can be calculated by evaluating $-\xi_\mu A^\mu$ in the horizon and subtracting its value at infinity,
\begin{equation}
    \phi = -(\xi_\mu A^\mu\big|_{r=r_+}-\xi_\mu A^\mu\big|_{r=\infty})\,,
\end{equation}
yielding 
\ba\label{phi}
    \phi &=& -a(r_+)-g \nonumber\\
    &=& g\left(\cos\left(e^{-\gamma}\left(\pi-2\arctan\frac{r_+}{n}\right)\right)-1\right)\nonumber\\
    &&+\frac{e}{2n} \sin\left(e^{-\gamma}\left(\pi-2\arctan\frac{r_+}{n}\right)\right)\,.
\ea
Upon using the electromagnetic duality \eqref{duality} we recover the corresponding magnetic potential
\ba\label{phim}
    \phi_m &=& \frac{e}{2n}\left(\cos\left(e^{-\gamma}\left(\pi-2\arctan\frac{r_+}{n}\right)\right)-1\right)\nonumber\\
    &&-g \sin\left(e^{-\gamma}\left(\pi-2\arctan\frac{r_+}{n}\right)\right)\,.
\ea

As usual in the framework of extended black hole thermodynamics \cite{Kubiznak:2016qmn}, 
we identify the cosmological constant with a dynamical pressure  
\begin{equation}
    P = -\frac{\Lambda}{8 \pi} = \frac{3}{8 \pi l^2}\,,
\end{equation}
allowing it to vary in the first law, and define the   thermodynamic volume as the corresponding conjugate quantity: 
\begin{equation}
    V = \left(\pdv{M}{P}\right)_{S,Q,N,\dots}\,.
\end{equation}

Finally, we calculate the Gibbs free energy {by evaluating the Euclidean action} 
\ba\label{eucAct}
{\cal I} &=& \tilde I + I_{\mbox{\tiny GH}} + I_{\mbox{\tiny C}} \\
    &=& \frac{1}{16\pi} \int_{M}\dd^4 x \sqrt{g}\left(R -2\Lambda\right)\nonumber\\
    &&{-\frac{1}{4\pi}\int \dd^4 x\sqrt{g}\tilde{\mathcal{L}}}\nonumber\\
    && +\frac{1}{8\pi} \int_{\partial M} \dd^3x \sqrt{h} \left(\mathcal{K} -\frac{2}{l} -\frac{l}{2} \mathcal{R}(h)\right),\nonumber
\ea
{where, apart from the bulk action $\tilde I$, given by \eqref{bulkAct} with 
${\cal L}$ replaced by $\tilde {\cal L}$ (see below), in the last line
we have also included} the Gibbons--Hawking term $I_{\mbox{\tiny GH}}$ and the AdS counterterm $I_{\mbox{\tiny C}}$ designed to cancel possible 
divergences \cite{Emparan:1999pm}. The Gibbs function is given by $\mathcal{G} = \mathcal{I}/\beta$, where $\beta$ is the inverse temperature and periodicity of the time Euclidean coordinate. 
{Note
that}{,} {in order to keep the metric and the vector potential real in the process of calculating
the action}{,} { one has to Wick rotate all of the following: the time coordinate, the NUT
parameter, and the electric and magnetic charge parameters. Interestingly, such a Wick rotated solution no longer solves the equations of motion derived from the non-linear electrodynamics Lagrangian ${\cal L}$, \eqref{NLELag}. In order} {for} { the Wick rotated solution} {to remain} {a solution of the corresponding equations of motion, one has to `replace' ${\cal L}$ with $\tilde{{\cal L}}$ obtained by `replacing' $P^2$ with $-P^2$ in \eqref{NLELag}, that is by considering\footnote{{This is the only example known to the authors in which the action of the theory has to be modified upon Wick rotation in order }{for} { the Euclidean solution}  {to remain} { a solution of the corresponding equations of motion.}} 
\be\label{NLELagWick}
    \tilde{\mathcal{L}} = -\frac{1}{2} \left({\cal S}\cosh\gamma - \sqrt{{\cal S}^2-{\cal P}^2}\sinh\gamma\right)\,.
\ee
Once the Euclidean action \eqref{eucAct} is calculated, we have to  Wick rotate back the NUT parameter and the electric and magnetic charges,  
upon which we obtain the following result:}
\begin{align}\label{GGG}
    \mathcal{G} =& \frac{m}{2} +\frac{eg}{2}\left(1-\cos\left(2e^{-\gamma}\left(\pi-2\arctan\frac{r_+}{n}\right)\right)\right)\nonumber\\
    &-\frac{e^2-4n^2g^2}{8n}\sin\left(2e^{-\gamma}\left(\pi-2\arctan\frac{r_+}{n}\right)\right)\nonumber\\
    &-\frac{r_+(3n^2+r_+^2)}{2l^2}\,.
\end{align}
In the limit $\gamma=0$, this expression straightforwardly reduces to that in the Maxwell theory, calculated in \cite{Ballon:2019uha}.

\subsection{Unconstrained thermodynamics}
Being in the grandcanonical ensemble, we have 
\ba\label{gibbs}
    \mathcal{G} &=& {\cal G}(T, \psi, \phi, Q_m^+, P)\nonumber\\
    &=&M - TS - \phi Q -\psi N\,,
\ea
and the Gibbs free energy satisfies 
\be
\delta G=-S\delta T-N\delta \psi -Q\delta \phi
+\phi_m \delta Q_m^++V\delta P\,. 
\ee

These relations can be used to verify the above expressions for $S, Q$ and $\phi_m$, as well as to find the remaining missing thermodynamic quantities $V$ and $N$. The thermodynamic volume reads   
\begin{equation}\label{volume}
    V = \frac{4}{3}\pi r_+^3\left(1+\frac{3n^2}{r_+^2}\right)\,,
\end{equation}
and is identical to the Maxwell case. The Misner charge $N$ can be calculated either from \eqref{gibbs}, or as 
\be
N=-\frac{\partial G}{\partial \psi}\Bigl|_{T,\phi, Q_m^+, P}\,. 
\ee
In either case the full expression for $N$ is not very illuminating and we omit it in this paper.

By construction the obtained thermodynamic quantities obey the full cohomogeneity first law:
\begin{equation}
\delta M=T\delta S+\psi \delta N + \phi \delta Q +\phi_{m}\delta Q_{m}^{+}+V \delta P\,,
\end{equation}
as well as the corresponding Smarr relation
\begin{equation}
M=2TS+\phi Q +\phi_{m}Q_{m}^{+}+2\psi N - 2 PV\,. 
\end{equation}

\subsection{Electric first law}
Similar to the Maxwell case \cite{Ballon:2019uha} let us finally consider the constrained thermodynamics, where one imposes an extra `regularity condition', 
\be
a(r+)=0\,, 
\ee
requiring that both the gauge potential $A$ \eqref{A} and the magnetic charge $Q_m^+$, \eqref{horizoncharges} vanish on the horizon. 
This allows one to eliminate the magnetic parameter according to 
\begin{equation}\label{eFirstLaw}
    g = -\frac{e}{2n}\tan\left(e^{-\gamma}\left(\pi-2\arctan\frac{r_+}{n}\right)\right).
\end{equation}
With this the electrostatic potential \eqref{phi} reads 
\begin{equation}
 \label{eqn:ElectricPotential}
    \phi = -g = \frac{e}{2n}\tan\left(e^{-\gamma}\left(\pi-2\arctan\frac{r_+}{n}\right)\right)\,,
\end{equation}
and the Gibbs free energy simplifies to 
\begin{align}
    \mathcal{G} =&{\cal G}(T, \psi, \phi, P)\nonumber\\
        =& \frac{m}{2}-\frac{r_+(3n^2+r_+^2)}{2l^2}\nonumber\\
    &-\frac{e}{2n}\tan\left(e^{-\gamma}\left(\pi-2\arctan\frac{r_+}{n}\right)\right)\,.
\end{align}
In this case, the thermodynamic volume takes the same form \eqref{volume}, while the Misner charge simplifies to  
\begin{align}
    N &= -\frac{4 \pi n^3}{r_+}\Big(1 + \frac{3(n^2-r_+^2)}{l^2} \nonumber\\
    &-e^2\Big(e^{-\gamma}\sec^2\left(e^{-\gamma}\left(\pi -2\arctan\frac{r_+}{n}\right)\right)\nonumber\\
&-\frac{r_+}{2n}\tan\left(e^{-\gamma}\left(\pi - 2 \arctan\frac{r_+}{n}\right)\right)\Big)\Big)\,.
\end{align}

With these simplifications we obtain the `electric first law' 
\begin{equation}
    \delta M = T \delta S + \phi \delta Q + \psi \delta N + V \delta P\,,
\end{equation}
accompanied by the following Smarr relation:
\begin{equation}
    M = 2(TS-VP + \psi N)+\phi Q\,.
\end{equation}

\section{Limit of vanishing NUT charge}\label{n0}

\subsection{Spherical solution}
Following \cite{Ballon:2019uha} let us now take the limit of a vanishing NUT charge. That is, we want to take the limit 
\be 
n \to 0\,,
\ee
while preserving both electric and magnetic charges at infinity:
\be
Q=e=\mbox{const.}\,,\quad 
Q_m=-2ng\equiv\hat g=\mbox{const.} 
\ee
Obviously, in order to keep the latter charge constant  as $n\to 0$ we have to have $g\to \infty$. 
The limit of the metric simply reads
\ba\label{RN}
ds^2&=&-f dt^2+\frac{dr^2}{f}+r^2(d\theta^2+\sin^2\!\theta d\phi^2)\,,\nonumber\\
f &=& 1 - \frac{2m}{r}+\frac{e^{-\gamma}(e^2 + \hat{g}^2)}{r^2}+\frac{r^2}{l^2}\,.
\ea

In order to take the limit of the gauge potential $A$, one has to first `renormalize' the gauge potential by adding a pure gauge, $A\to A+g dt$, upon which it is straightforward to take the limit and obtain 
\begin{equation}\label{RNA}
    A = -e^{-\gamma}\frac{e}{r} dt + \hat{g} \cos\theta d\phi\,.
\end{equation}

The solution given by \eqref{RN} and \eqref{RNA} represents the generalization of the (dyonic) charged AdS black hole to the case of the conformal electrodynamics. The thermodynamics of this solution follows from the thermodynamics of the Taub-NUT(-AdS) case discussed in the previous section. 
In particular, from \eqref{GGG} we obtain the following Gibbs free energy 
\begin{equation}
    \mathcal{G} = \frac{m}{2} - e^{-\gamma}\frac{e^2-\hat{g}^2}{2 r_+} - \frac{r_+^3}{2 l^2}\,.
\end{equation}
Since the Misner strings are no longer present, the horizon charges coincide with the asymptotic charges and the thermodynamics significantly simplifies.

\subsection{Self-dual solutions}
{Following \cite{Gibbons:2000mx}}{,} { let us finally briefly consider an interesting issue of Euclidean (anti)self-dual solutions in non-linear electrodynamics. For simplicity we limit ourselves to the self-dual spherical case; an extension to non-trivial NUT parameter is straightforward. Thus we consider a Wick-rotated solution \eqref{RN}, \eqref{RNA}, where we Wick rotate $t\to i\tau$ and $e\to ie$, to get
\ba\label{RNWick}
ds^2&=&f d\tau^2+\frac{dr^2}{f}+r^2(d\theta^2+\sin^2\!\theta d\phi^2)\,,\nonumber\\
A &=& e^{-\gamma}\frac{e}{r} d\tau + \hat{g} \cos\theta d\phi\,,\\
f &=& 1 - \frac{2m}{r}+\frac{e^{-\gamma}(\hat{g}^2-e^2)}{r^2}+\frac{r^2}{l^2}\,.\nonumber
\ea
As noted above, such a solution solves the equations of motion derived from $\tilde{\cal L}$, \eqref{NLELagWick} rather then ${\cal L}$, \eqref{NLELag}. 
}

The self-duality in non-linear electrodynamics now requires \cite{Gibbons:2000mx}:
\be\label{FE*}
F=*E\,.
\ee
In order to achieve this, we have to set 
\be\label{geqe}
\hat g =-e\
\ee
in (\ref{RNWick}). As is typical for self-dual solutions, despite having a nontrivial $F$, the corresponding electromagnetic tensor vanishes, $T^{\mu\nu}=0$, and the metric solves the vacuum with $\Lambda$ Einstein equations, see \cite{Gibbons:2000mx} for further properties.

Note also that the `traditional' self-duality condition 
\be\label{FF*}
F=*F 
\ee
is neither interesting to impose for non-linear electrodynamics (for example, the corresponding $T^{\mu\nu}$ does not necessarily vanish) nor does it automatically follow from \eqref{FE*}, as incorrectly stated in \cite{Gibbons:2000mx}. Our solution  \eqref{RNWick} with condition \eqref{geqe} provides an explicit example where \eqref{FE*} holds but \eqref{FF*} does not.

\section{Conclusion}\label{secConclusion}

We have obtained an exact axially symmetric black hole solution for Einstein gravity coupled to a maximally symmetric non-linear electromagnetic theory \eqref{NLELag}. 
While the black hole metric calculated above is identical to the Einstein--Maxwell case, except for the rescaling of the charges, a salient feature of the solution is that the electric and magnetic fields in this spacetime do not correspond to simple rescalings. As a consequence, it is in principle possible to detect any deviations from linear electrodynamics by placing charged test particles in a curved spacetime. Whether this observed behaviour for the solutions is exclusive to Taub-NUT class or is a general feature for other spacetimes in this theory remains to be explored. 

Moreover, we have extended the recent work on the thermodynamics of Lorentzian NUTs to this novel theory of electromagnetism, hence adding to the program of rehabilitating these spacetimes. Future work in these lines could aim towards studying the impact of non-linearity of the electromagnetic field on phase trasitions of the Taub-NUT solutions, as well as exploring the thermodynamics of other exact black hole solutions in this theory.
{When calculating the Euclidean action, we have observed a very interesting feature:  the Euclidean Wick-rotated solution no longer solves the equations of motion of the original non-linear electrodynamics Lagrangian \eqref{NLELag} but rather the modified Lagrangian \eqref{NLELagWick}, obtained from \eqref{NLELag} by replacing the invariant $P^2$ with $-P^2$. We expect this to be a general feature of any non-linear} theory of electrodynamics.

{Finally, we have discussed the Euclidean self-dual solution \eqref{RNWick}, \eqref{geqe}, emphasizing that the proper duality condition to impose is \eqref{FE*} instead of \eqref{FF*}, as customary in linear Maxwell theory.}

{\em Note added.}  After the work on the present manuscript was largely complete, it came to our attention that \cite{Flores-Alfonso:2020euz} had presented a similar solution to the one introduced here, albeit in a simpler spherically symmetric setup.

\section{Acknowledgements}
{We thank the anonymous referee for asking us to discuss self-dual solutions.}
    The work was supported in part by the Natural Sciences and Engineering Research Council of Canada.
Research at Perimeter Institute is supported in part by the Government of Canada through the Department of Innovation, Science and Economic Development Canada and by the Province of Ontario through the Ministry of Colleges and Universities. 
Perimeter Institute and the University of Waterloo are situated on the Haldimand Tract, land that was promised to the Haudenosaunee of the Six Nations of the Grand River, and is within the territory of the Neutral, Anishnawbe, and Haudenosaunee peoples.

\bibliographystyle{plain}  
\bibliography{referencesV2}

\end{document}